\documentclass{iau}

\usepackage{amsmath}
\usepackage{graphicx}
\usepackage{multirow}
\usepackage{orcidlink}

\defcitealias{GonzalezBolivar2023}{Paper~I}

\begin{document}

\lefttitle{Luis C. Bermúdez-Bustamante et al.}
\righttitle{Dust formation during the interaction of binary stars by common envelope.}

\journaltitle{Planetary Nebulae: a Universal Toolbox in the Era of Precision Astrophysics}
\jnlDoiYr{2023}
\doival{10.1017/xxxxx}
\volno{384}

\aopheadtitle{Proceedings IAU Symposium}
\editors{O. De Marco, A. Zijlstra, R. Szczerba, eds.}
 
\title{Dust formation during the interaction of binary stars by common envelope.
}

\author{Luis C. Bermúdez-Bustamante\orcidlink{0000-0002-3629-6259}$^{1,2}$, Orsola De Marco\orcidlink{0000-0002-1126-869X}$^{1,2}$, Lionel Siess\orcidlink{0000-0001-6008-1103}$^3$, Daniel J. Price\orcidlink{0000-0002-4716-4235}$^4$, Miguel González-Bolívar\orcidlink{0000-0002-5939-9269 }$^{1,2}$, Mike Y. M. Lau\orcidlink{0000-0002-6592-2036}$^{4,5,6}$, Chunliang Mu\orcidlink{0000-0003-1848-6507}$^{1,2}$, Ryosuke Hirai\orcidlink{0000-0002-8032-8174}$^{4,5}$, Ta\"issa Danilovich\orcidlink{0000-0002-1283-6038}$^{4,7}$, Mansi Kasliwal\orcidlink{0000-0002-5619-4938}$^8$}
\affiliation{$^1$School of Mathematical and Physical Sciences, Macquarie University, Balaclava Road, North Ryde, Sydney, NSW 2109, Australia\\
$^2$Astrophysics and Space Technologies Research Centre, Macquarie University, Balaclava Road, North Ryde, Sydney, NSW 2109, Australia\\
$^3$Institut d’Astronomie et d’Astrophysique, Université Libre de Bruxelles (ULB), CP 226, 1050 Brussels, Belgium\\
$^4$School of Physics and Astronomy, Monash University, Clayton, Victoria 3800, Australia\\
$^5$OzGrav: The ARC Centre of Excellence for Gravitational Wave Discovery, Australia\\
$^6$Heidelberger Institut f\"{u}r Theoretische Studien, Schloss-Wolfsbrunnenweg 35, 69118 Heidelberg, Germany\\
$^7$ARC Centre of Excellence for All Sky Astrophysics in 3 Dimensions (ASTRO 3D), Clayton 3800, Australia\\
$^8$Division of Physics, Mathematics, and Astronomy, California Institute of Technology, Pasadena, CA 91125, USA\\}

\begin{abstract}
We performed numerical simulations of the common envelope (CE) interaction between two intermediate-mass asymptotic giant branch (AGB) stars and their low-mass companions. For the first time, formation and growth of dust in the envelope is calculated explicitly. 
We find that the first dust grains appear as early as $\sim$1--3~yrs after the onset of the CE, and are smaller than grains formed later.
As the simulations progress, a high-opacity dusty shell forms, resulting in the CE photosphere being up to an order of magnitude larger than it would be without the inclusion of dust.
At the end of the simulations, the total dust yield is $\sim8.2\times10^{-3}$~\Msun ($\sim2.2\times10^{-2}$~\Msun) for a CE with a 1.7~\Msun (3.7~\Msun) AGB star. Dust formation does not substantially lead to more mass unbinding or substantially alter the orbital evolution.
\end{abstract}


\maketitle

\section{Introduction}
Unstable mass transfer from an expanding, more massive star to a compact star leads to a CE interaction  \citep{Ivanova2013}. The envelope of the giant eventually surrounds both the donor star's core and the companion. At the end of this phase, the envelope may be ejected and a binary system with a much smaller orbital separation or a merger is left behind.

The exact source(s) of energy driving the expansion and ejection of the envelope is not fully identified, and there are problems when considering only orbital energy and angular momentum transfer from the stars to the envelope \cite[but see][]{Valsan2023}, although the work done by recombination energy released in inner layers of the donor likely plays a major role \cite[e.g.,][]{Ivanova2015,Reichardt2020,Lau2022}.

The CE interaction leads to the formation of systems such as cataclysmic variables, X-ray binaries or progenitors of type Ia supernovae \citep{Paczynski1971,Ivanova2013,DeMarco2017}. It is also likely responsible for at least a fraction of intermediate luminosity transients \citep{Kasliwal2011}, such as luminous red novae \citep[e.g.,][]{Blagorodnova2017}, some of which produce dust \citep{Tylenda2011,Nicholls2013}.
From the above, dust might act as an additional driving mechanism, altering the appearance of transient objects by increasing their opacity.

Previous studies on dust in interacting binary stars were conducted in 1D \citep{Lu2013}, analyzed envelope thermodynamic properties without explicitly calculating dust formation \citep{Glanz2018,Reichardt2019}, calculated dust formation but in post-processing \citep{Iaconi2019b,Iaconi2020}, or focused on the phase before the common envelope \citep{Bermudez2020}.  

In \citet[][hereafter paper I]{GonzalezBolivar2023} we performed CE simulations with intermediate-mass AGB stars, using a simplified formulation by \citet{Bowen1988} to calculate the dust opacity. We find that dust does not significantly increase the mass loss, but it does increase the envelope opacity. Unfortunately, the Bowen formulation does not provide us with information about the properties of the dust, such as its mass or grains size. For this reason, we perform CE simulations with an explicit calculation of the dust formation process.

\section{Numerical Setup}
\label{sec:setup}

We use the same stellar models as in
\citetalias{GonzalezBolivar2023}, calculated with the 1D stellar evolution code of \mesa \cite[e.g.,][]{Paxton2011}. The first model is an AGB star, evolved from the main sequence, with a total mass of 1.7~\Msun (0.56~\Msun in the core), a radius of 260~\Rsun, undergoing its seventh thermal pulse. The second model is an AGB star with a total mass of 3.7~\Msun (0.72~\Msun in the core), a radius of 343~\Rsun, experiencing its fourth thermal pulse.
Both models feature a 0.6~\Msun companion star initially positioned at the critical distance for triggering Roche lobe overflow, i.e., at a distance of 550~\Rsun (637~\Rsun) from the 1.7~\Msun (3.7~\Msun) AGB star.

We use the smoothed particle hydrodynamics (SPH) code \phant \citep{Price2018} with $1.37\times10^6$ SPH particles as was done by \citet{GonzalezBolivar2022}.
We also use \phant's implementation of \mesa's OPAL/SCVH equation of state  tables \citep[as done by][]{Reichardt2020} to account for recombination energy.
We assume the formation of carbon-rich dust with a C/O ratio of 2.5, despite both AGB stars having a C/O ratio of 0.32 when our \mesa models ceased. The C/O ratio is not anticipated to impact the AGB convective structure.

We use the treatment described in \citet{Siess2022} for carbon dust nucleation and growth, which takes place in two steps: i) the formation of seed particles from the gas phase in the so-called nucleation stage, followed by ii) a growth phase where monomers (dust building blocks) accumulate on the seed particles to reach macroscopic dimensions. 
The treatment is based on the theory of moments presented by \citet[][and subsequent papers]{Gail1984}. 

\section{Dust properties and its impact on orbital and thermodynamics properties}
\label{sec:Results}



\subsection{Opacity of dust grains}

\begin{figure}[h]
    \centering
    \includegraphics[width=0.4\linewidth]{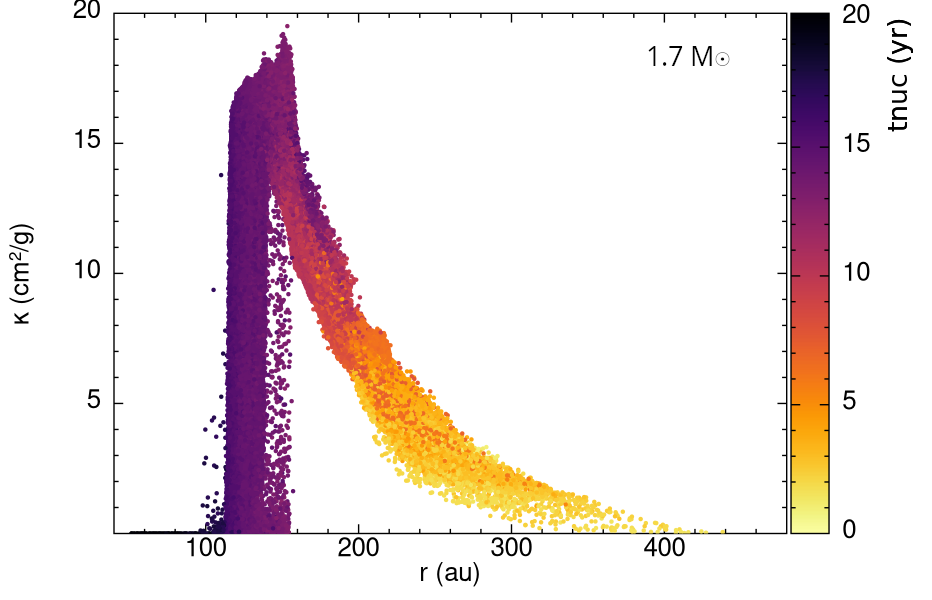}
    \includegraphics[width=0.4\linewidth]{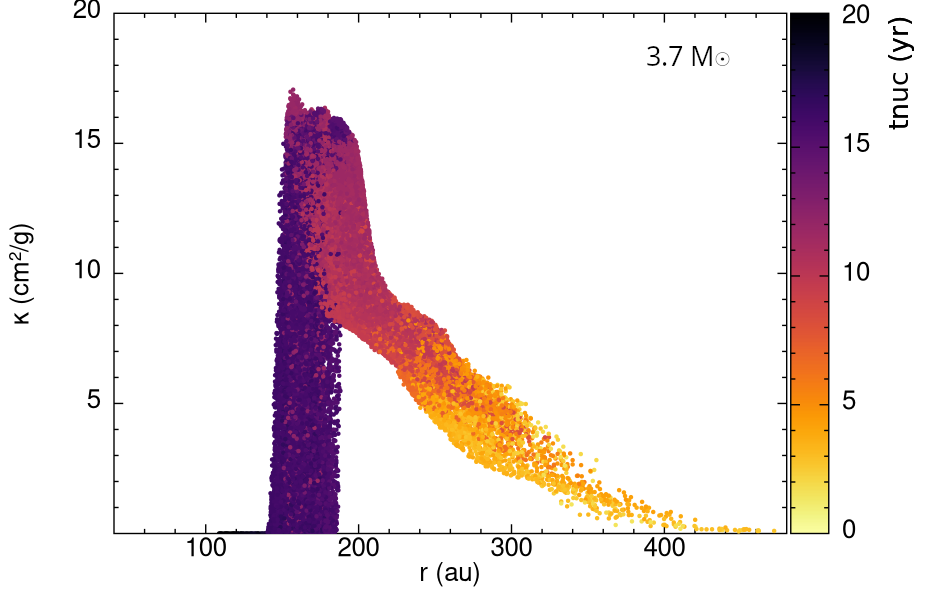}
    \caption{Dust opacity versus distance from the centre of mass for the 1.7~\msun (left) and 3.7~\msun (right) models after 20 years of simulation. The color bar indicates $t_{\rm nuc}$, the time in years after the start of the simulations, when dust nucleation occurred for each SPH particle.}
    \label{fig:opacity-versus-distance}
\end{figure}

Figure~\ref{fig:opacity-versus-distance} shows a snapshot\footnote{Movies corresponding to the figures are available at: \url{https://tinyurl.com/y455avdj}.} of the dust opacity as a function of distance from the center of mass (the color bar depicts nucleation time $t_{\rm nuc}$ as explained in the caption) at 20 years from the start of the simulation.
In the inner regions where there is no dust, the opacity is constant, at a user-defined value of 2$\times$10$^{-4}$~cm$^2$\,g$^{-1}$. This value is appropriate for recombined gas, but too low for the ionised regions where electron scattering opacities are of the order of 0.5~cm$^2$\,g$^{-1}$. However, this is not important for now as these inner hot regions do not experience dust formation, which happens $\sim$1--3~yrs from the start of the simulation, in a shell with radii of $\sim$10--20~au.
When dust forms, the opacity rapidly increases to values of $\sim$17--20~cm$^2$\,g$^{-1}$ and decreases to almost zero again at larger radii. The dust forming shell radius expands over time and at 20 years it is $\sim$130~au (160~au) from the 1.7~\Msun (3.7~\Msun) model. The dust forming shell thickness is also variable. 

\subsection{Dust size}
\label{ssec:dust_size}

\begin{figure}[h]
    \centering
    \includegraphics[width=0.4\linewidth]{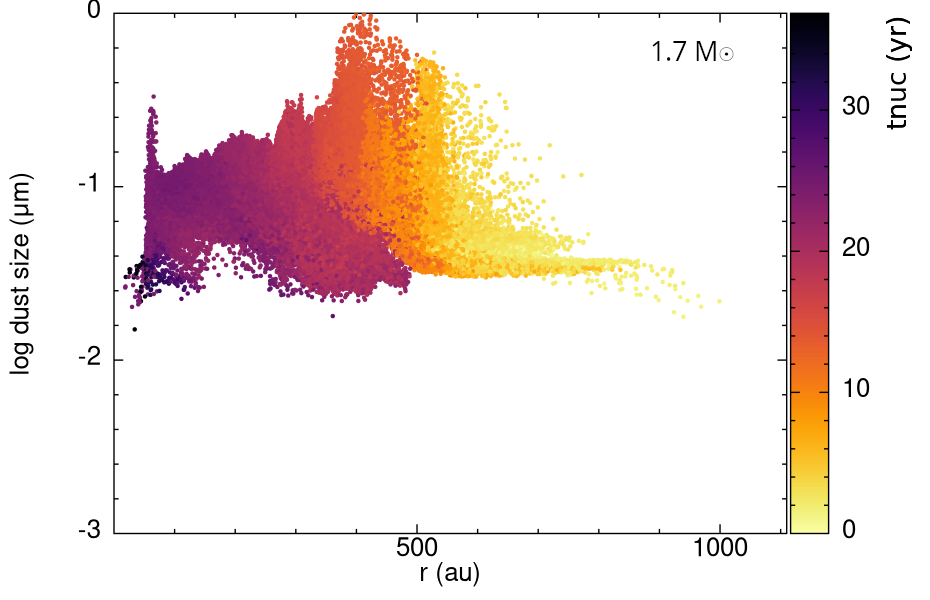}
    \includegraphics[width=0.4\linewidth]{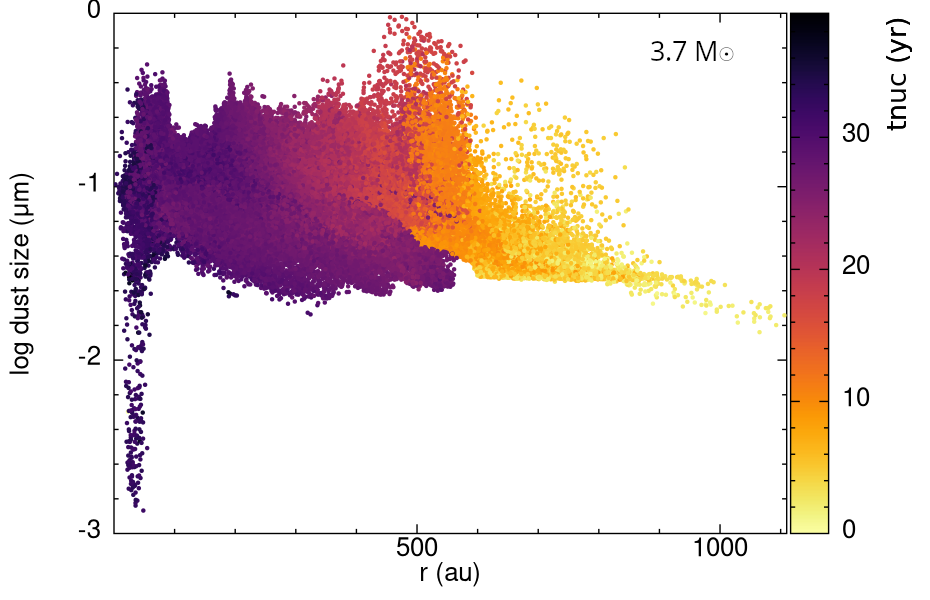}
    \caption{Average dust grains size versus distance for the 1.7~\msun (top) and 3.7~\msun (bottom) model, after 44 years of simulations. Same color bar as in Figure~\ref{fig:opacity-versus-distance}.} 
    \label{fig:median-size-distribution}
\end{figure}

Figure~\ref{fig:median-size-distribution} shows a snapshot of dust grains size as a function of distance (same color bar as in Figure~\ref{fig:opacity-versus-distance}). Early dust grains (yellow colour in Figure~\ref{fig:median-size-distribution}) have a narrow range in size around $\sim$0.02--0.04~$\mu$m, and late dust grains (darker colours) have a lager size between $\sim$0.03 and 1~$\mu$m. In the 1.7~\Msun (3.7~\Msun) model, the largest dust grains are formed at $\sim$12~yrs ($\sim$15~yrs).
Interestingly, at the very end of the 3.7~\Msun simulation ($\sim$44 years), dust formation resumes in a thin shell near the centre, but at a low rate.

Depending on $t_{\rm nuc}$, dust grains grew to different maximum sizes.
Our formalism does not include dust destruction, so once a grain stops growing, its size remains constant. However, what is plotted in Figure~\ref{fig:median-size-distribution} is the average grain size in a given SPH particle, which can contain many dust grains because it has a set total mass, e.g., $1.7\times 10^{27}$~g or $8.3\times 10^{-7}$~\Msun\ in the 1.7~\Msun\ simulation. As such, the average dust grain size in an SPH particle can decrease if new small grains are formed, even after previously-formed grains have become large.

\subsection{Dust mass}
\label{ssec:dust_mass}

\begin{figure}[h]
    \centering
    \includegraphics[width=0.4\linewidth]{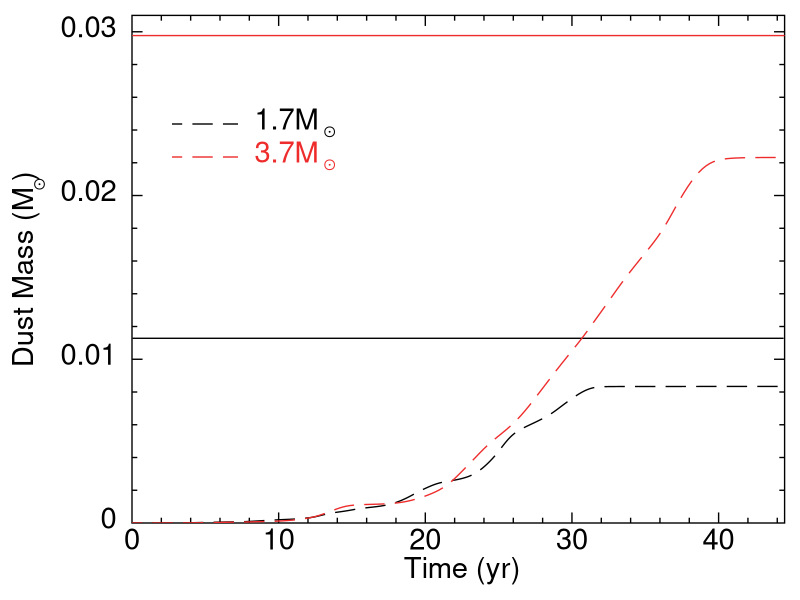}
    \caption{Dashed lines show the total dust mass over time in CE simulations. Horizontal lines represent the maximum carbon available in the envelope for 1.7~\Msun (black solid) and 3.7~\Msun (red solid) AGB stars.} 
    \label{fig:dust-mass-vs-time}
\end{figure}

Figure~\ref{fig:dust-mass-vs-time} shows that for the 1.7~\msun (3.7~\msun) model, the total dust mass increases steadily until it plateaus at $\sim$32~yrs ($\sim$38~yrs) with a value of $\sim$8.2$\times$10$^{-3}$~\msun ($\sim$2.2$\times$10$^{-2}$~\msun).
Although it is difficult to establish a pattern with only two stars, we see that the more massive the donor star, the longer it takes for the total amount of dust to plateau.

After $\sim$40~yrs, the 1.7~\Msun (3.7~\Msun) model has produced roughly four (ten) times more dust than during the first $\sim$20~yrs. The drastic increase and subsequent plateauing of dust mass is a consequence of a large fraction of the expanding envelope reaching temperatures below the condensation threshold. Because of that, nucleation is efficient (i.e., there is a large number of dust seeds) and the addition of monomers contribute to dust growth, producing the increase in the dust mass. 
However, with further gas expansion the nucleation rate decreases, fewer monomers are available and dust production is consequently reduced.

Compared to simulations by \citet{Iaconi2020} for carbon grains, the amount of dust formed in our 1.7~\Msun (3.7~\Msun) model is four (ten) times larger for a star twice (four times) as massive. In \citet{Iaconi2020}, the total dust mass plateaus at $\sim2.2\times10^{-3}$~\msun after $\sim$14~yrs for a 0.88~\Msun RGB star, which started off with a radius of 83~\Rsun. It is difficult to comment on whether these differences are due to the different type of stars, to the fact that \citet{Iaconi2020} post-processed the dust formation in their star, or to the nucleation formalism used.

\subsection{Orbital evolution, unbound mass and comparison with non-dusty simulations}
\label{ssec:orbital-evolution}

\begin{figure}[h]
    \centering
    \includegraphics[width=0.4\linewidth]{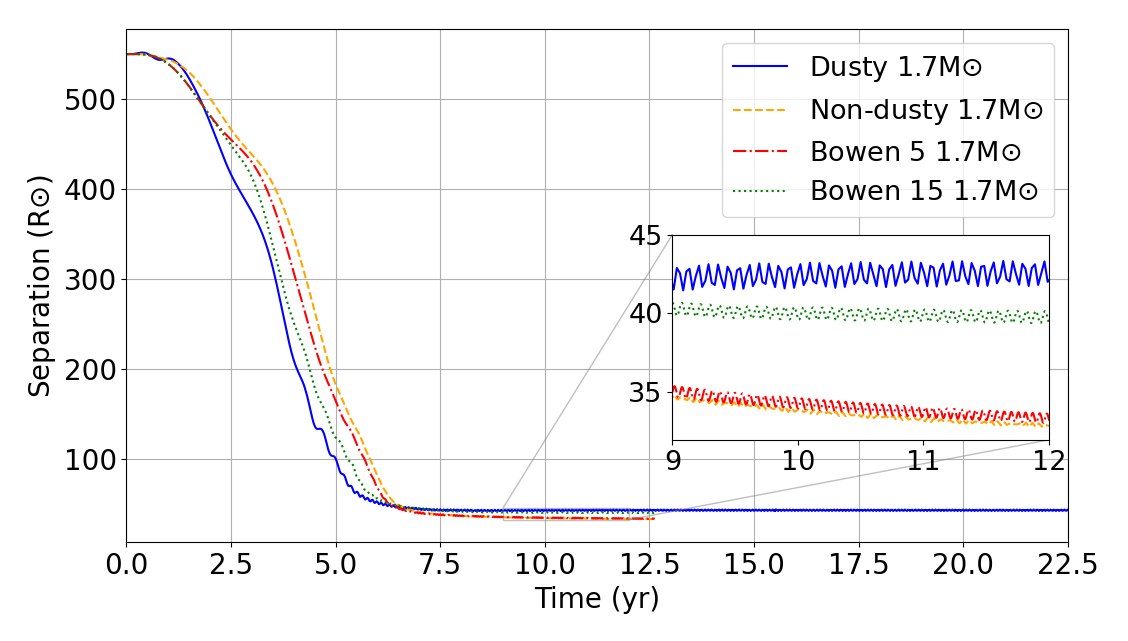}
    \includegraphics[width=0.4\linewidth]{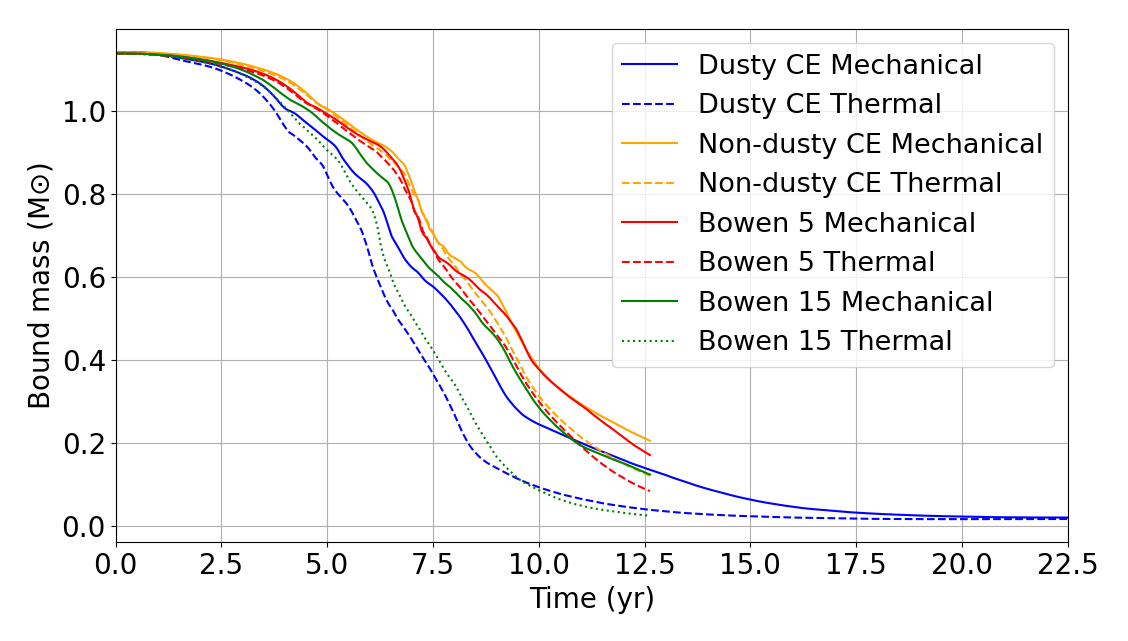}
    \caption{Left panel: orbital separation over time in CE simulations with a 1.7~\msun AGB star. Dust formation is explicitly calculated (blue lines), or assumed with a maximum dust opacity of $5~\mathrm{cm^2\,g^{-1}}$ (red lines) or $15~\mathrm{cm^2\,g^{-1}}$ (green lines) as in \citetalias{GonzalezBolivar2023}. Also shown is a CE simulation without dust (yellow lines). The inset displays the orbital separation at the end of the in-spiral phase. Right panel: Mechanical bound mass (solid lines) and thermal bound mass (dashed lines) over time.}
    \label{fig:separation_massunbound}
\end{figure}

For all simulations, the dynamic in-spiral phase ends after $\sim$7.5 yr (Figure~\ref{fig:separation_massunbound}, left panel). 
The orbital separation decreases faster in models where dust opacity is calculated either explicitly (blue line), or using the Bowen opacity approximation (green and red lines), than in the model with no dust (yellow line). Radiative forces have a significant impact on accelerating the envelope, which interact through gravitational friction with the stars.
On the other hand, the final (plateau) separation is similar, being $\sim$42~\Rsun (blue line) and $\sim$33~\Rsun
(yellow line), although the latter is still decreasing somewhat.

In the right panel of Figure~\ref{fig:separation_massunbound}, ``mechanical" bound mass (solid lines) is the mass of SPH particles whose sum of kinetic and gravitational potential energy is negative, while ``thermal" bound mass (dashed lines) is the mass of SPH particles whose sum of kinetic, gravitational and thermal energy (not including recombination energy) is less than zero.
The bound mass decreases faster in models with explicit dust formation (blue lines) than in models (i) without dust (yellow lines) or (ii) with a maximum Bowen opacity of 5~$\mathrm{cm}^2 \mathrm{g}^{-1}$ (red lines); however, when the latter increases to $15~\mathrm{cm}^2 \mathrm{g}^{-1}$ (green lines), the evolution of the bound mass is similar. 
High-opacity dust is located too far in the wind\footnote{See movies of dust opacity versus distance at \url{https://tinyurl.com/y455avdj}.} to generate effective driving on a dynamical timescale required to aid with envelope ejection, because at those distances, the radiative flux is greatly diluted and the material is mostly unbound already.
We conclude that dust formation does not lead to substantially more mass unbinding, although $\sim10^{-2}$~\msun of dust is produced. 

\section{The photospheric size}
\label{sec:photo}

\begin{figure}[h]
    \centering
    \includegraphics[width=0.4\linewidth]{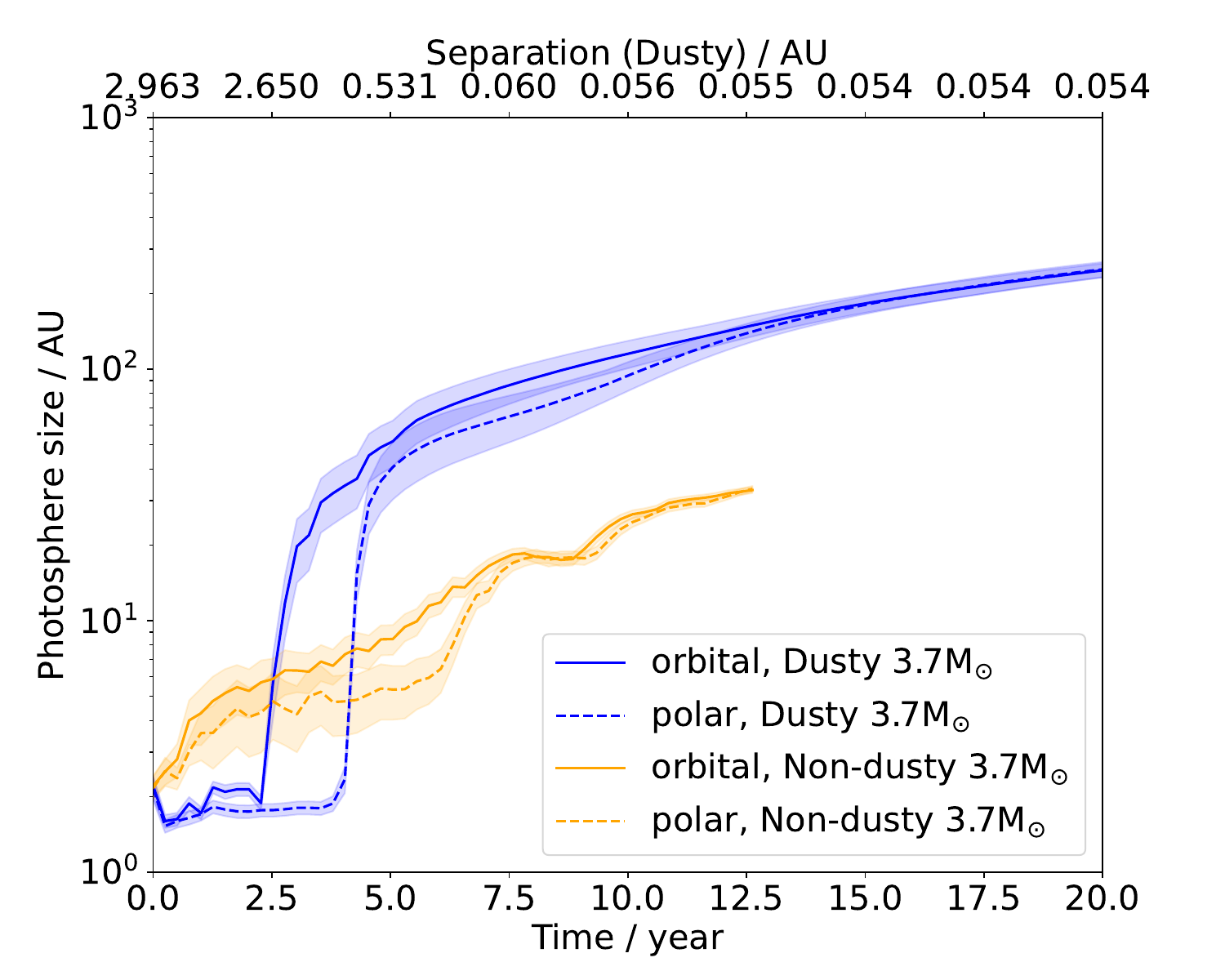}
    \caption{Photosphere size over time in the orbital (solid lines) and the ``polar" plane (dashed lines) for the 3.7~\Msun model with and without dust (blue and orange, respectively). Shaded areas depict uncertainty.}
    \label{fig:photo-size-vs-sep}
\end{figure}

Figure~\ref{fig:photo-size-vs-sep} shows the average radius, in the polar and orbital directions, of the stellar surface where the optical depth, calculated using the Planck-mean opacity, becomes 1 (i.e. the photosphere).
We note that after $\sim$5~yrs the size of the photosphere is an order of magnitude larger when dust is included. This is expected, since dust opacity will be high enough for the expanding envelope to become optically thick.
We also note that in the dusty model, the equatorial photosphere grows approximately one year before the polar photosphere, effectively indicating that there is a “hole" at the poles where an observer would see deeper into the object. However, this hole is filled rapidly and the size of the photosphere in the two orthogonal directions becomes very comparable, indicating an approximately spherical object.

\section{Summary}
\label{sec:Conclusions}

We have carried out simulations of CE interactions of two intermediate-mass AGB stars (1.7 and 3.7~\Msun) with a 0.6~\msun companion. To test the dust grains properties, simulations include a self-consistent treatment of dust formation with a C/O ratio of 2.5 (by number). Our main conclusions are:

\begin{enumerate}
\item Dust formation starts $\sim$1--3~yrs after the onset of the CE and leads to the formation of a dusty shell that moves outward. At $\sim$20~yrs the shell has a thickness of $\sim40$~au and it is located at $\sim130$~au ($\sim160$~au) in the 1.7~\Msun (3.7~\Msun) model.

\item Dust begins as tiny seeds and grows via monomer addition. The growth halts once the dust moves away along with the expanding envelope. Dust grains formed at $\sim$1--3~yrs are smaller ($\sim$0.02--0.04~$\mu$m) than dust grains formed after $\sim$10~yrs ($\sim$0.03--1~$\mu$m). 

\item The amount of dust rises and plateaus just below the theoretical maximum, yielding $\sim8.2\times10^{-3}$~\Msun ($\sim2.2\times10^{-2}$~\Msun) for the 1.7~\Msun (3.7~\Msun) model.

\item High-opacity dust is located too far in the wind to generate effective driving on a dynamical timescale required to aid with envelope ejection. Because of that, although $\sim10^{-2}$~\msun of dust is produced, it does not lead to substantially more mass unbinding.

\item Dust greatly impacts the optical appearance of the CE. At $\sim$12~yrs, the photosphere is an order of magnitude larger if dust is included.
\end{enumerate}

Movies corresponding to the figures are available at: \url{https://tinyurl.com/y455avdj}.\\

OD, LS, MK and LB acknowledge support through the Australian Research Council Discovery Program grant DP210101094. LS is senior research associates from F.R.S.- FNRS (Belgium).  TD is supported in part by the Australian Research Council through a Discovery Early Career Researcher Award (DE230100183). This work was supported in part by Oracle Cloud credits and related resources provided by Oracle for Research. This work was performed in part on the OzSTAR national facility at Swinburne University of Technology. The OzSTAR program receives funding in part from the Astronomy National Collaborative Research Infrastructure Strategy (NCRIS) allocation provided by the Australian Government, and from the Victorian Higher Education State Investment Fund (VHESIF) provided by the Victorian Government. Some of the simulations were undertaken with the assistance of resources and services from the National Computational Infrastructure (NCI), which is supported by the Australian Government. This research was supported in part by the Australian Research Council Centre of Excellence for All Sky Astrophysics in 3 Dimensions (ASTRO 3D), through project number CE170100013.


\end{document}